# COVID Lessons: Was there any way to reduce the negative effect of COVID-19 on the United States economy?

Mohammadreza Mahmoudi

*Department of Economics, Northern Illinois University, Dekalb, USA.*

Email: mmahmoudi@niu.edu

**Abstract**

This paper aims to study the economic impact of COVID-19. To do that, in the first step, I showed that the adjusted SEQIER model, which is a generalization form of SEIR model, is a good fit to the real COVID-induced daily death data in a way that it could capture the nonlinearities of the data very well. Then, I used this model with extra parameters to evaluate the economic effect of COVID-19 through job market. The results show that there was a simple strategy that US government could implemented in order to reduce the negative effect of COVID-19. Because of that the answer to the paper's title is yes. If lockdown policies consider the heterogenous characteristics of population and impose more restrictions on old people and control the interactions between them and the rest of population the devastating impact of COVID-19 on people lives and US economy reduced dramatically. Specifically, based on this paper's results, this strategy could reduce the death rate and GDP loss of the United States 0.03 percent and 2 percent respectively. By comparing these results with actual data which show death rate and GDP loss 0.1 percent and 3.5 percent respectively, we could figure out that death rate reduction is 0.07 percent which means for the same percent of GDP loss executing optimal targeted policy could save 2/3 lives. Approximately, 378,000 persons dead because of COVID-19 during 2020, hence reducing death rate to 0.03 percent means saving around 280,000 lives, which is huge.

# 1. Introduction

COVID-19 as the one of the main challenges of human being in new era changed our lives dramatically, in the way that many scientists believe that our era could divided into pre-COVID and post-COVID periods. COVID-19 appeared first time in Wuhan, China but with astonishing speed spread all over the world. The reason behind creation of COVID-19 is mysterious yet, however, it takes many lives and makes huge damage in economy of countries. From late November 2019, which China announced the first case of COVID-19 in Wuhan up until December 2021 278 million positive COVID-19 cases and 5.39 million COVID-induced death have recorded worldwide. In addition, the IMF report [1] shows in advanced economies over 2020–22 the cumulative per capita income losses are expected to be 11 percent of 2019 per capita GDP. Whilst, in emerging and developing countries (excluding China) losses are equivalent to 20 percent of 2019 per capita GDP. Although, public vaccination was available at the beginning of 2021, too many undeveloped countries are falling behind, and economic inequality is worsening.

In the United States as the first country with most COVID-19 cases and death toll, 51.8 million positive COVID-19 cases and 814,000 COVID-induced death have recorded as of December 2021. For getting comprehensive understanding about COVID-19 outbreaks in the first step I summarized all the important COVID-19 related events in the United States which have happened from December 2019 up until December 2021 in the Table 1.

*Table 1. Timeline of the COVID-19 pandemic in the United States*

| | Date | Description |
|---|---|---|
| **2020** | Jan 9 | WHO announced unknown pneumonia virus (Wuhan-Hu-1) in Wuhan, China |
| | Jan 11 | China reported its first death |
| | Jan 20 | CDC confirmed the first U.S. coronavirus Case |
| | Jan 21 | Chinese Scientist Confirms coronavirus could transmit by human |
| | Jan 22 | Trump tweets coronavirus is not a threat |
| | Jan 30 | The first case of person-to-person transmission was detected in Chicago |
| | Jan 31 | WHO declared a global health emergency |
| | Feb 2 | The U.S. restricted travel from China |
| | Feb 3 | The U.S. reported public health emergency |
| | Feb 6 | The U.S. reported the first coronavirus-induced death |
| | Feb 11 | The disease the virus causes was named, COVID-19. |
| | Feb 24 | Trump tweets the coronavirus is very much under control in the U.S. |
| | Feb 25 | The CDC declared the status of COVID-19 is pandemic |
| | Mar 13 | Trump administration declared a national emergency |
| | Mar 15 | The CDC issued guidance recommended no gathering of 50 or more people for an eight-week period. |
| | Mar 21 | Illinois's authorities announced stay at home order |
| | Mar 26 | Senate passed CARES Act |
| | **As of March 31, 3,170 deaths, 164,620 confirmed cases, and 1.07 million tests were reported in the U.S.** ||
| | Apr 16 | "Gating Criteria"[2] proposed as a way to reopen the economy |
| | Apr | IRS issued the first COVID-19 stimulus checks |
| | **As of April 30, 60,966 total deaths, 1.04 million confirmed cases, and 6.25 million tests were reported in the U.S.** ||
| | May 9 | Saliva-based diagnostic test allowed for at-home use |
| | May 12 | Fauci testified that death toll likely undercounted |
| | May 26 | The George Floyd protests begun in Minneapolis. |
| | May 28 | The U.S. COVID-19 Deaths Pass the 100,000 |
| | **As of May 31, 103,781 total deaths, 1.77 million confirmed cases, and about 14 million tests were reported in the U.S.** ||
| | June 10 | The U.S. COVID-19 Cases reached 2 million |
| | June 22 | A study suggested 80% of COVID-19 cases in March went undetected |
| | **As of June 30, 126,140 total deaths, 2.59 million confirmed cases, and about 30 million tests were reported in the U.S.** ||
| | July 2 | States authorities reversed reopening plans |
| | July 9 | WHO declared COVID-19 can be airborne |
| | July 13 | More than 5 million Americans lost health insurance. |
| | July 16 | The U.S. reported new record of daily COVID-19 cases |
| | July 28 | The CDC called for schools reopening |
| | Aug 17 | COVID-19 is the third-leading cause of death in the U.S. |
| | Aug 28 | The first known COVID-19 reinfection case reported in the U.S. |
| | Aug 30 | The U.S. COVID-19 positive cases passed 6 million. |
| | Sept 22 | The U.S. COVID-induced death toll surpassed 200,000. |
| | Sept 25 | The U.S. COVID-19 positive cases passed 7 million. |
| | Sept 28 | Global COVID-induced death toll reached 1 million. |
| | Sept 29 | First presidential debate held |
| | Oct 2 | Trump tested positive for COVID-19, and went to hospital |
| | Oct 19 | Global COVID-19 positive cases surpassed 40 million |
| | Oct 22 | The FDA approved Remdesivir as the first COVID-19 drug |
| | Oct 30 | The U.S. COVID-19 positive cases passed 9 million |
| | Nov 3 | The U.S. presidential election held |
| | Nov 4 | Unprecedented positive cases in 1 day, 100,000, recorded in the U.S. |
| | Nov 27 | The U.S. COVID-19 positive cases passed 13 million |
| | Dec 11 | The FDA approved a vaccine by Pfizer. |
| | Dec 14 | First vaccine given in US as roll-out begun |
| | Dec 17 | The U.S. COVID-induced death toll surpassed 300,000. |

|  | Dec 27 | The U.S. COVID-19 positive cases passed 19 million cases |
|---|---|---|
|  | Dec 30 | A confirmed case of the new variant from the UK was reported in Colorado. |
|  | Dec | The IRS issued the second COVID-19 stimulus check |
| 2021 | Jan 14 | More than 1 million persons were vaccinated |
|  | Jan 15 | A confirmed case of the new variant from the UK was reported in Illinois. |
|  | Jan 18 | The U.S COVID-induced death toll passed 400,000. |
|  | Jan 22 | The U.S. COVID-19 positive cases passed 25 million cases |
|  | Jan 25 | First U.S. positive case of Brazil variant of coronavirus reported in Minnesota. |
|  | Jan 26 | First U.S. positive case of South African variant of coronavirus reported in South Carolina. |
|  | Feb 20 | The U.S. COVID-19 positive cases passed 28 million cases |
|  | Feb 22 | The U.S. COVID-induced death toll passed 500,000 |
|  | Feb 23 | More than 1,880 cases of the UK variant were reported in 45 states |
|  | Mar 1 | More than 51 million people have received one or more Covid-19 vaccine doses |
|  | Mar 8 | The CDC declared that fully vaccinated people can gather indoors without masks. |
|  | Mar 13 | The U.S. passed 100 million vaccinations administered. |
|  | Mar 11 | American Rescue Plan signed and the third COVID-19 stimulus check was issued by IRS |
|  | Apr 21 | U.S. passed 200 million vaccinations administered. |
|  | June 1 | The Delta variant, became the dominant variant in the U.S. |
|  | Oct 7 | A CDC study revealed more than 140,000 U.S. children under the age of 18 years lost their parents |
|  | Nov 26 | World Health Organization classified a new variant, Omicron |

Note: this table summarized the main COVID-19 related events in the United States that have happened from December 2019 up until December 2021. [3]

In order to control COVID-19 spread, different countries implemented distinct policies including creating emergency committees, allocating fund to expand testing capacity, restricting national and international travels, closing schools and business, and shelter in place orders. One of the main social distancing policies that adopted by federal and states authorities in different states in the United States is shelter in place order. Based on shelter-in-place order, all residents must stay at home homes for all but essential activities including buy food or medicine, exercise, care for others, and essential travels. California is the first state in the United States, which implemented shelter in place order as the COVID-controlling policy on March 19, 2020. As of April 20, 2020, 40 states and the District of Columbia had implemented shelter in place orders. The timeline of COVID-controlling policies is summarized in the Figure 1.

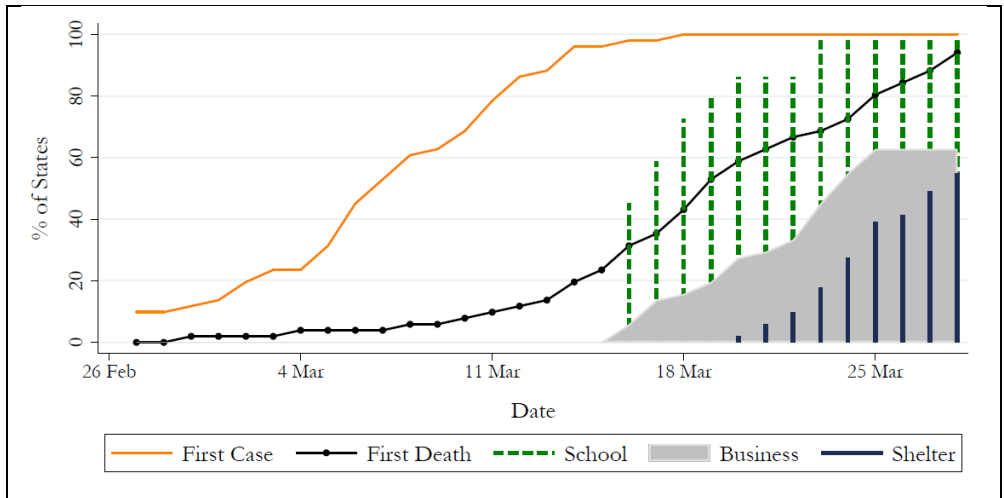

Figure 1. COVID-controlling policies in the United States

Note: This figure plots the COVID-controlling policies which were implemented by federal and local authorities in the United States. The black line represents the percentage of states with a first COVID-induced death. The orange line shows the percentage of states that have at least one Covid-19 confirmed case. The grey shaded area represents business closures. The dashed spikes indicate shelter-in-place policies. Source: Brzezinski et al. (2020)

By analyzing the lockdown policies, we can figure out they treat all people equally and they ignore the heterogeneous characteristics of population. However, COVID-19 have affected age group differently. Based on CDC studies, risk of infection, hospitalization, and death of COVID-19 are different for various age group. This fact is shown in Table 2. The reference age group is the 18-29 year-old category, because this group has the largest cumulative number of COVID-19 positive cases compared to other age groups. For example, compared to the 18-29 year-old individuals, hospitalization and death for people older than 85 years are 15 and 570 times higher, respectively.

Table 2. Risk for COVID-19 Infection, hospitalization, and death by age

|  | 0-4 years old | 5-7 years old | 18-29 years old | 30-39 years old | 40-49 years old | 50-64 years old | 65-74 years old | 75-84 years old | 85+ years old |
|---|---|---|---|---|---|---|---|---|---|
| **Cases** | <1x | 1x | Reference group | 1x | 1x | 1x | 1x | 1x | 1x |
| **Hospitalization** | <1x | <1x | Reference group | 2x | 2x | 4x | 5x | 9x | 15x |
| **Death** | <1x | <1x | Reference group | 4x | 10x | 30x | 90x | 220x | 570x |

Note: This table shows risk for COVID-19 infection, hospitalization, and death by age group compared to 18-29 years old age group. Compared to the 18-29 year-old individuals, hospitalization and death for people older than 85 years are 15 and 570 times higher, respectively. [4]

Real data also confirmed the contents of Table 2. For example, number of positive cases by age group is shown in the Figure 2. Based on the data from 34,430,784 cases up until October 4, 2021, the highest positive cases belong to 18-39 year-old individuals. In fact, 38 percent of the positive cases are young (18-39 years old), 34 percent are middle aged (40-64 years old), and only 12 percent of positive cases are old (more than 65 years old).

*Figure 2.Number of COVID-19 positive cases by age group*

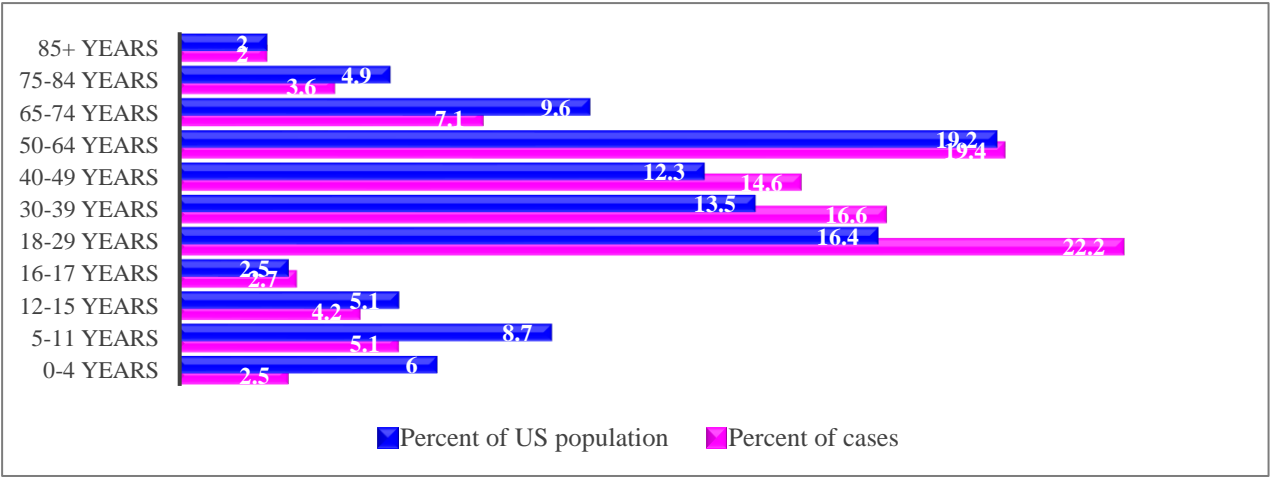

Note: This figure shows the number of positive cases by age group up until October 04, 2021. Based on the figure, 38 percent of the positive cases are young (18-39 years old), 34 percent are middle aged (40-64 years old), and only 12 percent of positive cases are old (more than 65 years old). [6]

Moreover, by analyzing the data of deaths by age group we find that the 65 years and older group accounted for the highest death rate. As it is described in Figure 3, only 2.2 percent of COVID-induced death was from young adults (18-39 years old), around 20 percent of COVID-induced death was from middle-aged adults (40-64 years old), and around 80 percent of COVID-induced death was from adults 65 years and older.

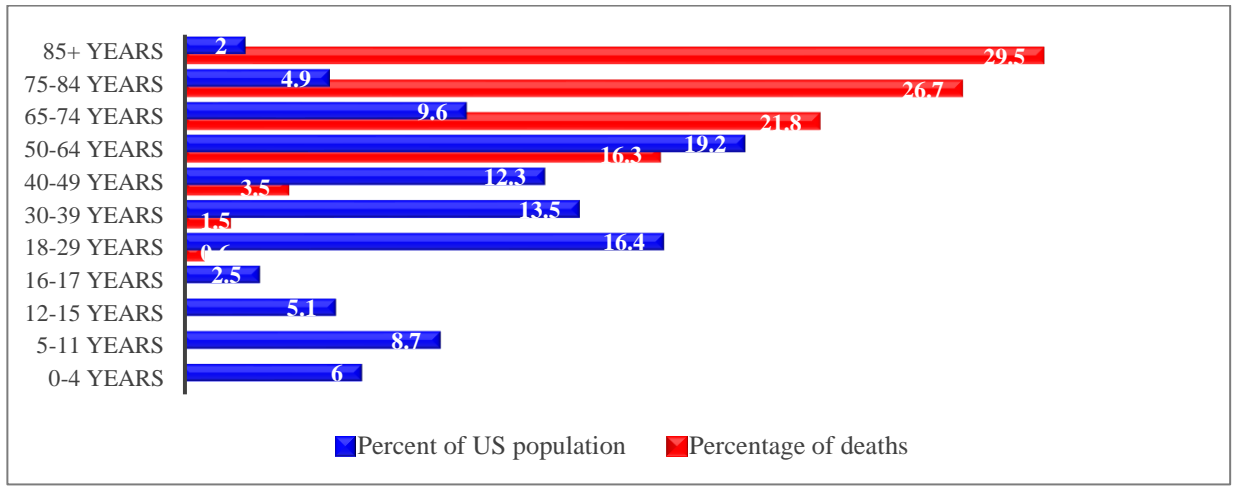

Figure 3. Number of COVID-induced death by age group

Note: This figure shows the number of COVID-induced death by age group up until October 04, 2021. Based on the figure, 2.2 percent of COVID-induced death was from young adults (18-39 years old), around 20 percent of COVID-induced death was from middle-aged adults (40-64 years old), and around 80 percent of COVID-induced death was from adults 65 years and older. [6]

These statistics motivate me to study COVID-19 modeling in the case of heterogenous population. By putting the age statistics of COVID-19 pandemic together we could conclude that 1) we have heterogenous population, and 2) the risk of getting infected and dying are different for each age group. Therefore, every optimal policy should consider the heterogeneity aspect of COVID-19. For example, one of the first conclusions that we could get from above data analysis was that: Since the majority of positive cases belong to 18-29 year old and the majority of death belongs to those who are 65 years and older, the optimal policy should reduce the interaction within and between these groups.

In this paper, I use an adjusted SEQIHR model to study the economic effect of COVID-19 for United States. First, I introduce some important papers, which focus on compartmental models to examine COVID-19 and its economic impacts. Then, I use data and run the SEQIHR model to find the best fitted model to COVID-induced daily death data of United States as well as analyze the economic effect of COVID-19 by using multi-risk SEQIHR model to address the heterogenous feature of COVID-19 and evaluate COVID-controlling policies. Finally, I discuss main results and provide conclusion.

## 2. Literature review

John Graunt (1620–1674) is the first scientist who studied infectious disease data. After him Daniel Bernoulli (1700–1782) used a mathematical epidemiology model to evaluate the effect of inoculation on smallpox. In 1906, W.H. Hamer introduced the basic idea of compartmental models. He asserted that the spread of infection should rely on the susceptible and infected populations. During that time Sir Ronald Ross introduced a simple compartmental model in order to analyze the transmission of malaria between mosquitoes and humans. He also mentioned the concept of basic reproduction number for the first time in his work (Brauer, Castillo-Chavez, and Feng (2019)). Moreover, Kermack and McKendrick (1927), Kermack and McKendrick (1932), and Kermack and McKendrick (1933) studied the transmission of infection using the basic compartmental models profoundly. Later, their models reformulated and used by Diekmann, Heesterbeek, and Metz (1995) to studied different disease outbreaks containing the SARS pandemic of 2002–2003, the H5N1 influenza outbreak in 2005, the H1N1 influenza epidemic in 2009, and the Ebola epidemic in 2014.

Compartmental models as the mathematical modeling of infectious diseases divided population into several compartments. Based on defined compartments, we have different compartmental models such as SIR, SIS, SIRD, SEIR, and so on (Brauer, Castillo-Chavez, and Feng (2019)). For example, in SIR model, which is known as the simplest compartmental model, the population divided into three categories: susceptible, infected, and recovered. Another famous compartmental model is SEIR model which have extra compartments named exposed group. There are many papers which employed a compartmental model to explore an outbreak (e.g. Blackwood and Childs (2018); GILL ; Gibson and Renshaw (1998); Worden and Porco (2017); Hethcote (2000); Berger, Herkenhoff, and Mongey (2020); Arias et al. (2021); Arias et al. (2021); Yan and Zou (2006)). Silal et al. (2016) compared different compartmental models in epidemiology. They argued that the compartmental models generated distinct outcomes but when fitted to the data, these results were robust to the model structure. Also, indicating the purpose of the models is very vital in the case of using complex compartmental models.

In this paper, I used an adjusted SEQIHR model to analyze COVID-19 pandemic. For the first time Gumel et al. (2004) introduced SEQIHR model to study the SARS epidemic of 2002–2003. They divided population into six categories of susceptible, exposed, infected, quarantined, isolated/hospitalized, and recovered/dead to examine the effect of isolation and quarantine to control SARS outbreaks in Toronto, Beijing, Singapore, and Hong Kong. Their results indicated that the main strategy to control SARS is reducing the contact rate between susceptible and infectious groups by isolating infected individuals. Also, governments by mixing optimal isolation policies and highly effective testing programs could eradicate SARS. In addition, Siriprapaiwan, Moore, and Koonprasert (2018) did the similar approach work with updated data set and got the same results. There are many papers which utilized the compartmental model to analyze COVID-19 pandemic. In the following I summarized the results of some of the main papers.

Bertozzi et al. (2020) argued that due to high fluctuation of reproduction number over time and by location, inaccuracy of COVID-19 death and positive cases data, and the difficulties of evaluating nonpharmaceutical interventions, modeling and forecasting the COVID-19 outbreaks is very challenging process. In addition, Roda et al. (2020) introduced the non-identifiability in model calibrations using the confirmed-case data as the main resource of inaccurate COVID-19 predictions. They argued that more complex compartmental models may not lead to more reliable model. COVID, Team, and Hay (2020) used COVID-19 data from 1 February 2020 to 21 September 2020 and a SEIR model to evaluate possible scenarios and the effect of COVID-controlling policies. They anticipated 511,373 (469,578–578,347) COVID-induced death for United States by 28 February 2021. Moreover, they predicted universal mask use reduce death toll by 129,574 (85,284–170,867) from September 22, 2020 to the end of February 2021.

Moreover, Rahmandad, Lim, and Sterman (2021) developed a behavioral dynamic and used a hierarchical Bayesian approach to estimate the model parameters with a panel dataset of 91 countries. They estimated cumulative cases and deaths through 30 October 2020 to be 8.5 and 1.4 times greater than official reports, leading an overall infection fatality rate (IFR) of 0.48% and showed modest policy interventions and behavior change could decrease cumulative cases by 18%. Wang et al. (2020) proposed a Bayesian framework to estimate epidemiological patterns and evaluate the effect of intervention strategies on the

COVID-19 outbreak. In addition, Atkeson (2021b) presented a behavioral epidemiological model of COVID-19 pandemic based on 2020 available data in the US and UK. By considering seasonal variation of transmission rate and pandemic fatigue, he developed a model which fitted to the real data of US and UK very well, even his model could capture the complex dynamic of COVID-19 which created by new variant of pandemic. Atkeson, Kopecky, and Zha (2020a) used a SIR model to estimate and forecast COVID-19 pandemic. The results showed that they predicted the second wave of COVID-19 in the United States correctly. Also, they claimed that it is no clear whether earlier lockdown policies would have decreased the long-term cumulative COVID-induced death. Furthermore, Sargent and Stachurski (2015) in their comprehensive book "Quantitative Economics with Julia", specifically section V: Modeling in Continuous Time, provided the Julia code for deterministic and stochastic COVID-19 modeling.

There are some papers which focus on the limitations of COVID-19 modeling. For instance, Fernández-Villaverde and Jones (2020) used COVID-induced death data in the U.S. states and different countries to estimate a standard COVID-19 epidemiological model. They simulated a SIRD model to evaluate $R_0$ and predicted baseline mortality rate (IFR) based on different scenarios and recognized a substantial uncertainty about IFR. They estimated $R_0 = 2.7$ for New York City which social distancing could decrease it to 0.5. In addition, Atkeson (2020a) ran an SIR model and showed how estimation of COVID-19 fatality rate is difficult in the first month of pandemic because of the data limitation. Moreover, he indicated that the model with a high initial number of COVID-19 positive cases and a low death rate and the same model with a low initial number of COVID-19 positive cases and a high death rate provide the same predictions regarding the pandemic evolution. There are other recent papers that used compartmental models in order to analyze COVID-19 pandemic (e.g. Atkeson (2020b); Atkeson et al. (2020); Atkeson, Kopecky, and Zha (2021); Acemoglu, Makhdoumi, et al. (2020); Acemoglu et al. (2021); Chernozhukov, Kasahara, and Schrimpf (2021); Willebrand (2021)).

In addition, there are many economists who study the economic impact of COVID-19, among them I address several main papers in the following. One of the economists who study COVID-19 modeling and its effect on economy profoundly is Atkeson (2021a). He used a model of private and public behavior to study the dynamic of COVID-19 and answer to the question how we could mitigate the effect of COVID-19 pandemic on public health. He concluded that the combination of quick non-pharmaceutical interventions and fast development of pharmaceutical technologies could save 300,000 lives approximately. Also, Atkeson, Kopecky, and Zha (2020b) detected four facts regarding the effect of non-pharmaceutical interventions on COVID-19 outbreak. 1- 20-30 days after each region experienced 25 cumulative deaths, the growth rate of COVID-19 daily death reduced dramatically. 2- After this period, the daily death growth rate constant around zero. 3- In the first 10 days of pandemic the standard deviation of daily death growth rate decreased dramatically. 4- Based on epidemiological models, these three facts interpreted as the reduction of COVID-19 reproduction number and transmission rate. Also, they argued that ignoring these four facts leads to exaggerate the importance of non-pharmaceutical policies. Avery et al. (2020) provided the critical reviews regarding the coronavirus pandemic models from economist point of view. They claimed that it is not clear whether COVID-19 models help policy makers to implement effective COVID-controlling policies in the first months of pandemic, even in some cases like UK it leaded the UK government responded to the spread of the pandemic with delay. They argued that profound data set leads to more accurate COVID modelling. (Also, see Brodeur et al. (2021)).

Stock (2020) provided a framework for economists which used SIR model to understand the impacts of social distancing and containment policies on the development of pandemic and economy. He got the asymptomatic rate using random sampling of the population. To evaluate the economics of shutdown policies he found the most efficient policies to achieve a given transmission rate, $\beta$, then determined the path of $\beta$ in order to compute the trades off between economic cost and lives loss. Moreover, there are several papers which studied the effect of COVID-controlling policies by considering heterogenous population. In the following I summarized the results of several of the main papers, then discussed the main difference between my approach and theirs.

[Kaplan, Moll, and Violante (2020)](#) integrated an extended SIR model of COVID-19 into a macroeconomic model realistic parameters and job-related and sectoral heterogeneity. They maintained that the economic welfare costs of COVID-19 are large and heterogenous for all combinations of health and economic policies, because of that governments should consider the lives loss costs as well as the fraction of population who endure the upset of economic costs in order to design the COVID-controlling policies. [Brotherhood et al. (2020)](#) investigated the importance of the age structure in the COVID-19 outbreak by using an expanded SIR model. They found that imposing more restrictions on young people will extend the duration of pandemic, as well as testing and quarantines save lives even when they only targeted young people. Also, government can raise the welfare of young and old age groups by mixing different policies. Old people can protect themselves during the pandemic. [Rampini (2020)](#) scrutinized a sequential framework to elevate COVID-19 interventions policies by considering a heterogeneous population. He divided population into to group: young people who less affected by COVID-19 and work more and old people who more affected by COVID-19 and work less, then figured out lifting lockdown for young people first and the old people later leads to reduction in demand for health care system, death rate, and economy cost of lockdown.

In addition, [Crossley, Fisher, and Low (2021)](#) examined the UK labor market shocks, and the way individuals and household responded to the first wave of COVID-19 based on the available data regarding first two waves of COVID-19 outbreak. Their results reveled that people with fragile employment condition including those belong to minority ethnic groups, and under 30 years old are affected by labor market shocks significantly. Also, 50 percent of individuals had experienced reduction in their income by 10 percent, and this reduction is very severe for poor households. [Arnon, Ricco, and Smetters (2020)](#) evaluated the early lockdown policies using an epidemiological-econometric model. They created daily measures of contact rate and employment and estimate main parameters of the model using an event study framework.

The results showed that 1- non-pharmaceutical interventions reduced death toll by 30 percent and saved 33,000 lives, and decline employment by 15 percent, around 3 million jobs in the first three months on COVID-19 outbreak. However, there is small evidence that indicated these policies reduced contact rates nationwide over the same period. 2- Stay at home order is more effective strategy than shutdown policy which targeted businesses. 3- A strict lockdown policies at the first months of pandemic could improve epidemiological and economic consequences in the meantime.

[Acemoglu, Chernozhukov, et al. (2020)](#) developed a multi-risk SIR model where each compartment divided into three age groups, young, middle-aged, and old in order to capture the heterogenous aspect of COVID-19 outbreak. The results showed that the optimal targeted policies which imposed strict restrictions on old people outperform uniform policies significantly. Specifically, optimal semi–targeted or fully-targeted policies reduce mortality from 1.83% to 0.71%, hence, save 2.7 million lives compared to uniform policies. Another paper which focusses on multi-risk SIR model is written by [Gollier (2020)](#). Specifically, he studies the effect of age-specific confinement and PCR testing policies on incomes and death. He claimed that investing 15% of annual GDP to confine 90% of population for 4 months or confine 30% of population for 5 months could eliminate COVID-19. However, he asserted that paper's results depend on uncertain epidemiological, economics, social, and policy parameters. [Glover et al. (2020)](#) paper provided more comprehensive approach regarding heterogenous characteristics of COVID-19 with respect to previous papers. They made a model to determine economic activity and pandemic development. They divided population by young and old age groups, by basic and luxury economic activity, and by health status. Moreover, they analyzed the optimal economic alleviation policy in the case that government can reallocate taxes and transfers. Their results show optimal reallocation and alleviation policies interact, and more modest lockdown policies are optimal when reallocation makes tax distortions. In addition, they claimed that if vaccine available in the first months of 2020, harder but shorter lockdown will be best strategy to control COVID-19. In addition, I study a bunch of related papers which examine the economic impact of COVID-19 including [Wu and Olson (2020)](#); [Boppart et al. (2020)](#); [Favero, Ichino, and Rustichini (2020)](#);

Friedman et al. (2020); Heesterbeek and Roberts (2007); Zhang et al. (2021); Cajner et al. (2020) ; Baqaee et al. (2020); Baqaee and Farhi (2021); Baqaee and Farhi (2020); Baker et al. (2020); Hall, Jones, and Klenow (2020); McKibbin and Fernando (2020); Farboodi, Jarosch, and Shimer (2021); Guerrieri et al. (2020); Brinca, Duarte, and Faria-e-Castro (2020); Bloom, Kuhn, and Prettner (2020); La Torre, Malik, and Marsiglio (2020); Bourne (2021); Chen et al. (2021); Barua (2020); McKibbin and Fernando (2021); Bloom, Fletcher, and Yeh (2021); Eichenbaum, Rebelo, and Trabandt (2021).

## 3. Methodology

I used the adjusted SEQIHR model to model COVID-19 outbreaks. This model is a generalization of the SEIR model of an infectious disease in which a population is divided into six categories of susceptible (S), exposed (E), infected (I), quarantined (Q), isolated/hospitalized (H), and recovered/dead (R). The susceptible group represents uninfected people that can be infected by contacting with infected or exposed individuals. The exposed group refers to individuals who are potentially infective but cannot transmit infection. The infected group denotes symptomatic individuals who can transmit infection. The quarantined group are those COVID-infected individuals without clinical symptoms who are separated from others, while the isolated group refers to the separation of COVID-infected individuals with clinical symptoms from others. Moreover, recovered group represents the people who are recovered or dead from the disease.

I added several components to the SEQIHR model that reflected some key epidemiological properties of the COVID-19 pandemic.

- I considered the natural rate of birth and death and their effects on the COVID-19 dynamic.
- In order to consider endogenous response of policy makers and individuals during the pandemic, I considered some of the main parameters of the model, like contact rate and basic reproduction number over time.
- I added the rate of vaccination of susceptible individuals ($v$), because at the end of 2020 public vaccination was available in most states.

The parameters that I used in my model are summarized in Table 3.

Table 3. Parameters for the SEQIJR model

| Parameters | Description |
|---|---|
| N | Total population |
| S | Susceptible group |
| E | Exposed group |
| I | Infected group |
| Q | Quarantined group |
| H | Isolated/hospitalized group |
| R | Recovered group |
| D | Dead individuals |
| $\Pi$ | Natural birth rate |
| $\mu$ | Natural death rate |
| $\upsilon$ | Rate of vaccination of susceptible individuals |
| $\beta$ | Effective contact rate |
| $\varepsilon_E$ | Modification parameter related to infection from exposed group |
| $\varepsilon_Q$ | Modification parameter related to infection from quarantined group |
| $\varepsilon_H$ | Modification parameter related to infection from hospitalized group |
| $s_R$ | Rate of recovered people who are endanger again (people are not fully immune against COVID-19 reinfection) |
| $\gamma_E$ | Rate of quarantine of exposed asymptomatic individuals |
| $\gamma_I$ | Rate of isolation of infectious symptomatic individuals |
| $r_I$ | Rate of recovery of infectious symptomatic individuals |
| $r_H$ | Rate of recovery of hospitalized individuals |
| $r_Q$ | Rate of recovery of quarantined individuals |
| $\sigma_E$ | Rate of development of symptoms in asymptomatic individuals |
| $\sigma_Q$ | Rate of hospitalization of quarantined individuals |
| $d_I$ | Rate of COVID-induced death for symptomatic individuals |
| $d_H$ | Rate of COVID-induced death for isolated individuals |

Note: This table summarized all the required parameters to run SEQIHR model. It should be noted that the rates are per day.

## 3.1  SEQIHR Model Assumptions

- The susceptible group (S(t)) expands by natural birth ($\Pi$) and recovered people who are endangered again with rate $s_R$ (this is because recovered people are not fully immune against COVID-19 reinfection). Moreover, it shrinks by natural death at rate µ, recovered people because of vaccination at rate ν, and contact by infected people (symptomatic, asymptomatic, quarantined, or isolated) who go to other compartments with transmission rate $\beta, \beta\varepsilon_E, \beta\varepsilon_Q, and \beta\varepsilon_H$. In fact, quarantine is not perfect but reduces the contact rate by a factor $\varepsilon_Q$, and isolation is imperfect, but decreases the contact rate by $\varepsilon_H$ .

- The exposed group (E(t)) increases by new infections of the susceptible group. It is decreased by people becoming infective at rate $\sigma_E$, being quarantined at a rate $\gamma_E$, and dying at rate µ.

- The infected group (I(t)) expands by developing clinical symptoms among asymptomatic individuals at rate $\sigma_E$. It reduces by natural death at rate μ, COVID-induced death at rate $d_I$, hospitalization at rate $\gamma_I$, and recovery at rate $r_I$.
- The quarantined group (Q(t)) generates by susceptible and exposed individuals who are quarantined at rate $\varepsilon_Q$, and $\gamma_E$ respectively. It shrinks by recovery at rate $r_Q$, hospitalization at rate $\sigma_Q$, and natural death at rate μ.
- The hospitalized group (H(t)) increases by infected and quarantined individuals who are isolated at hospitals at rate $\gamma_I$ and $\sigma_Q$ respectively, and diminished by natural death at rate μ, COVID-induced death at rate $d_H$, and recovery at rate $r_H$

Based on these assumptions the graph of the adjusted SEQIHR model is depicted in Figure 4:

*Figure 4.The Adjusted SEQIHR Model*

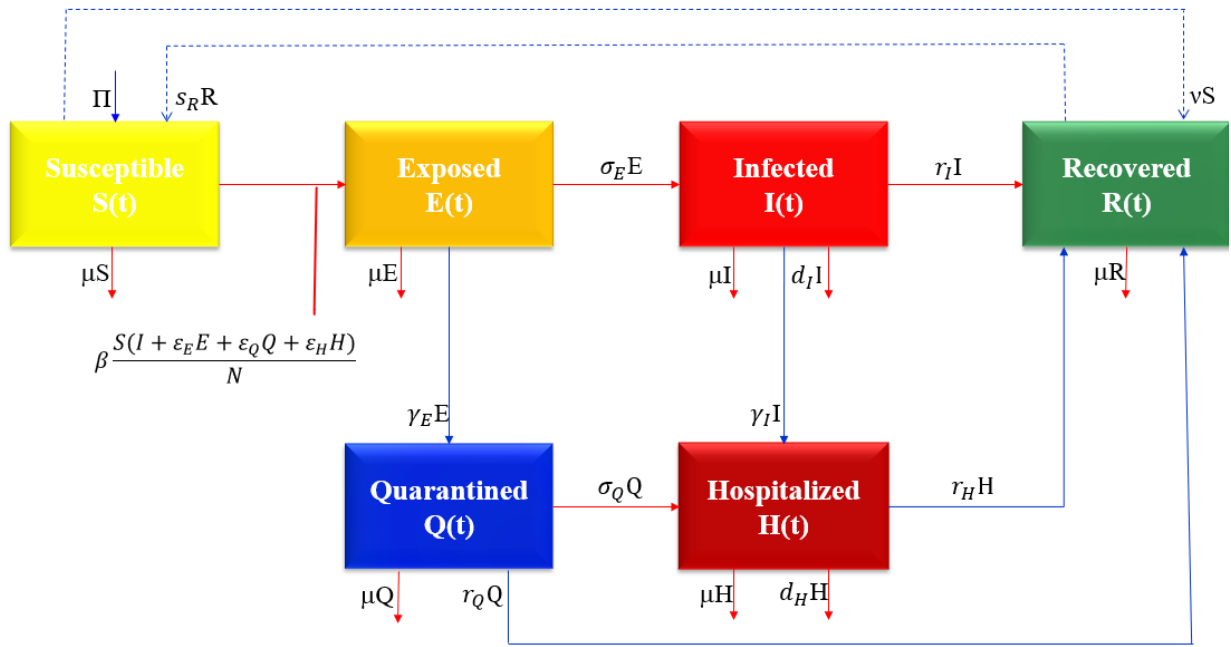

Note: Based on the SEQIHR model assumptions I depict this plot. The SEQIHR model has six compartments susceptible (S), exposed (E), infected (I), quarantined (Q), isolated/hospitalized (H), and recovered/dead (R). The input and output of each compartment determine by the SEQIHR model assumptions.

Moreover, the model assumptions lead to the following system of ordinary differential equations:

$$\frac{dS}{dt} = \Pi - \frac{SL}{N} - M_S S + s_R R \qquad (1)$$

$$\frac{dE}{dt} = \frac{SL}{N} - M_E E \qquad (2)$$

$$\frac{dI}{dt} = \sigma_E E - M_I I \qquad (3)$$

$$\frac{dQ}{dt} = \gamma_E E - M_Q Q \qquad (4)$$

$$\frac{dH}{dt} = \gamma_I I + \sigma_Q Q - M_H H \qquad (5)$$

$$\frac{dR}{dt} = \nu S + r_I I + r_H H - M_R R \qquad (6)$$

Since $N(t) = S(t) + E(t) + I(t) + Q(t) + H(t) + R(t)$, $\frac{dN}{dt} = \Pi - \mu N - d_I I - d_H H$.

In addition, death rate induced by the COVID-19 is as follows:

$$\frac{dD}{dt} = d_I I + d_H H \qquad (7)$$

Where, $L = \beta(I + \varepsilon_E E + \varepsilon_Q Q + \varepsilon_H H)$, is contact rate of susceptible individuals with infected people (symptomatic, asymptomatic, quarantined, or isolated).

Moreover, $M_S = (\nu + \mu)$, $M_E = (\gamma_E + \sigma_E + \mu)$, $M_I = (\gamma_I + d_I + r_I + \mu)$, $M_Q = (r_Q + \sigma_Q + \mu)$, $M_H = (d_H + r_H + \mu)$ and $M_R = (\mu + s_R)$ are outflow rate coefficients for susceptible (S), exposed (E), infected (I), quarantined (Q), isolated/hospitalized (H), and recovered/dead (R) compartments.

At the steady state we have $\frac{dS}{dt} = \frac{dE}{dt} = \frac{dI}{dt} = \frac{dQ}{dt} = \frac{dH}{dt} = \frac{dR}{dt} = \frac{dN}{dt} = 0$, Therefore, there are two equilibriums for the system of questions.

i. **Disease-free equilibrium,** which is not a useful solution for our problem.

$$\frac{dS}{dt} = 0 \rightarrow S_0^* = \frac{\Pi}{M_S} \tag{8}$$

$$\frac{dE}{dt} = 0 \rightarrow E_0^* = p \tag{9}$$

$$\frac{dI}{dt} = 0 \rightarrow I_0^* = 0 \tag{10}$$

$$\frac{dQ}{dt} = 0 \rightarrow Q_0^* = 0 \tag{11}$$

$$\frac{dH}{dt} = 0 \rightarrow H_0^* = 0 \tag{12}$$

$$\frac{dR}{dt} = 0 \rightarrow R_0^* = \frac{vS^*}{\mu+s_R} = \frac{v\Pi}{M_R M_S} \tag{13}$$

$$\frac{dN}{dt} = 0 \rightarrow N_0^* = \frac{\Pi}{\mu} \tag{14}$$

ii. **Pandemic equilibrium,** which is the targeted solution for our system of equations.

$$\frac{dS}{dt} = 0 \rightarrow S^* = \frac{1}{M_S}(\Pi - \alpha_S I^*) \tag{15}$$

$$\frac{dE}{dt} = 0 \rightarrow E^* = \alpha_E I^* \tag{16}$$

$$\frac{dI}{dt} = 0 \rightarrow I^* = \frac{\Pi(\mu\alpha_I - M_S\alpha_S)}{\alpha_S(\mu\alpha_I - M_S\alpha_N)} \tag{17}$$

$$\frac{dQ}{dt} = 0 \rightarrow Q^* = \alpha_Q I^* \tag{18}$$

$$\frac{dH}{dt} = 0 \rightarrow H^* = \alpha_H I^* \tag{19}$$

$$\frac{dR}{dt} = 0 \rightarrow R^* = \frac{v\Pi}{\mu M_S} - \alpha_R I^* \tag{20}$$

$$\frac{dN}{dt} = 0 \rightarrow N^* = \frac{1}{\mu}(\Pi - \alpha_N I^*) \tag{21}$$

Where, $\alpha_E = \frac{M_I}{\sigma_E}$, $\alpha_S = M_E \alpha_E$, $\alpha_I = \beta(1 + \varepsilon_E \alpha_E + \varepsilon_Q \alpha_Q + \varepsilon_H \alpha_H)$, $\alpha_Q = \frac{\gamma_E}{M_Q} \alpha_E$, $\alpha_H = \frac{\gamma_I + \sigma_Q \alpha_Q}{M_H}$, $\alpha_R = \frac{1}{\mu}(\frac{v}{M_S}\alpha_S - r_I - r_H \alpha_H)$, and $\alpha_N = d_I + d_H \alpha_H$

**Proof:** First, we solve the system of equations for $S^*, E^*, Q^*, H^*, R^*,$ and $N^*$ as a function of $I^*$ to get the above equations. Then, by substituting the results into equation ( 2 ), we get $\alpha_S I^{*2}(\mu \alpha_I - M_S \alpha_N) - \Pi(\mu \alpha_I - M_S \alpha_S) = 0$. By solving this equation we get two values for $I^*$, $I^* = 0$, and $I^* = \frac{\Pi(\mu \alpha_I - M_S \alpha_S)}{\alpha_S(\mu \alpha_I - M_S \alpha_N)}$, which is equal to equation ( 17 ). Hence, the proof is complete. ∎

## 3.2 Reproduction Number

The basic reproduction number $(R_0)$ is one of the critical epidemiology metrics that measures the transmissibility of infectious agents and can be computed using the Lyapunov's first method. Based on the epidemiology literature, when the transmission epidemic speed is high like COVID-19, $R_0 > 1$, hence an epidemic can develop from a small number of infected people. Otherwise, when $R_0 < 1$, an epidemic cannot develop. In this thesis, I computed the COVID-19 basic reproduction number, $R_0$, for a situation in which we do not have controlling policies, and the control reproduction number, $R_c$, for a situation in which we have COVID-19 controlling policies. By using Lyapunov's first method, we could compute $R_0$ and $R_c$ for the SEQIHR model as follows:

First, we linearized pandemic equilibrium equations as $\frac{dX}{dt} = J^* X$

Where $X = (E, S, Q, I, H, R)^T - X = (E^*, S^*, Q^*, I^*, H^*, R^*)^T$ and $J^*$ is Jacobian matrix at pandemic equilibrium.

$$J^* = J(E^*, S^*, Q^*, I^*, H^*, R^*)$$

$$= \begin{bmatrix} J_{11} & J_{12} & J_{13} & J_{14} & J_{15} & J_{16} \\ -J_{11} - M_S & -J_{12} - M_E & -J_{13} & -J_{14} & -J_{15} & -J_{16} \\ 0 & \gamma_E & -M_Q & 0 & 0 & 0 \\ 0 & \sigma_E & 0 & -M_I & 0 & 0 \\ 0 & 0 & \sigma_Q & \gamma_I & M_H & 0 \\ \nu & 0 & 0 & r_I & r_H & -\mu \end{bmatrix} \quad (22)$$

Where,

$J_{11} = -\frac{L}{N} + \frac{LS}{N^2} - M_S$, $J_{12} = -\frac{\beta \varepsilon_E S}{N} + \frac{LS}{N^2}$, $J_{13} = -\frac{\beta \varepsilon_Q S}{N} + \frac{LS}{N^2}$, $J_{14} = -\frac{\beta S}{N} + \frac{LS}{N^2}$, $J_{15} = -\frac{\beta \varepsilon_H S}{N} + \frac{LS}{N^2}$, and $J_{16} = \frac{LS}{N^2}$.

$$\begin{aligned} det(J^*) = \{&\mu M_E M_I M_Q M_H LN + \mu M_I M_Q M_H M_S LS + \mu \gamma_E M_I M_H M_S LS + \mu \sigma_E M_Q M_H M_S LS \\ &+ \sigma_E r_I M_Q M_H M_S LS + \mu \gamma_E \sigma_Q M_I M_S LS + \mu \gamma_I \sigma_E \sigma_Q M_Q M_S LS \\ &+ \gamma_E \sigma_Q r_H M_I M_S LS + \gamma_I \sigma_E r_H M_Q M_S LS + \mu M_E M_I M_Q M_H M_S N^2 \\ &- (\mu M_E M_I M_Q M_H LS + \nu M_E M_I M_Q M_H LS + \mu \beta \sigma_E M_H M_Q M_S NS \\ &+ \mu \beta \varepsilon_E M_I M_H M_Q M_S NS + \mu \beta \varepsilon_Q \gamma_E M_I M_H M_S NS + \mu \beta \varepsilon_H \gamma_E \sigma_Q M_I M_S NS \\ &+ \mu \beta \varepsilon_H \gamma_I \sigma_E M_Q M_S NS)\}/N^2 \end{aligned} \quad (23)$$

$$det(J^*) = \frac{M_E M_I M_Q M_H M_S}{N \alpha_S} \{\mu \alpha_S N + \frac{\mu \alpha_S L}{M_S} + \frac{LS}{N}(r_I + r_H \alpha_H + \mu(1 + \alpha_E + \alpha_Q + \alpha_H))\}(1 - R_C) \quad (24)$$

Therefore, $R_C$ equal to

$$R_C = \frac{S}{N} \frac{\mu \alpha_L N + \alpha_S L}{\mu \alpha_S N + \frac{\mu \alpha_L L}{M_S} + \frac{LS}{N}(r_I + r_H \alpha_H + \mu(1 + \alpha_E + \alpha_Q + \alpha_H))} \quad (25)$$

## 3.3 SEQIHR Model Estimation

At time zero $S(0) > 0, E(0) > 0, and\ I(0) = Q(0) = H(0) = R(0) = D(0) = 0$. Moreover, the quantities of parameters are mentioned in Table 4.

Table 4.The related quantities of the SEQIHR model's parameters

| Parameters | Description | Quantities (Mean/Median) |
|---|---|---|
| $\Pi$ | Natural birth rate | The general fertility rate for US is 55.8 births per 1,000 women aged 15–44 in 2020.[7] |
| $\mu$ | Natural death rate | The natural death rate for US is 7.37 per 1000 people in 2020. [8] |
| N | Rate of vaccination of susceptible individuals | At the end of 2020 only 0.01% of the US population had been fully vaccinated. [9] |
| $\varepsilon_E$ | Modification parameter related to infection from exposed group | I set $\varepsilon_E = 0$, based on Gumel et al. (2004). |
| $\varepsilon_Q$ | Modification parameter related to infection from quarantined group | I set $\varepsilon_Q = 0$, based on Gumel et al. (2004). |
| $\varepsilon_H$ | Modification parameter related to infection from hospitalized group | I set $\varepsilon_H = 0.8$, based on Gumel et al. (2004). |
| $s_R$ | Rate of recovered people who are endanger again (people are not fully immune against COVID-19 reinfection)/ vaccine breakthrough infection | The vaccine is 95 percent effective. The data showed between January 17 and August 21, 2021,1 in 5,000 fully vaccinated people experienced a breakthrough infection. More recently, this rate is 1 in 100 fully vaccinated people. [10] |
| $\gamma_E$ | Rate of quarantine of exposed asymptomatic individuals | Mean time that infectious person should be quarantine is 2 weeks. [11] |
| $\gamma_I$ | Rate of isolation of infectious symptomatic individuals | Median number of days from symptom group to hospitalized group is 5 days. [11] |
| $r_I$ | Rate of recovery of infectious symptomatic individuals | Approximately 98% of COVID-19 infectious individual recovered. Wang et al. (2020) |
| $r_H$ | Rate of recovery of hospitalized individuals | Approximately 80% of patients hospitalized with COVID-19, and 60% of those admitted to ICUs, survive. [11] |
| $r_Q$ | Rate of recovery of quarantined individuals | On average a COVID patient needs 2 weeks to be recover. [11] |
| $\sigma_E$ | Rate of development of symptoms in asymptomatic individuals | Mean time from exposure group to symptom infected group is 6 days. [11] |
| $\sigma_Q$ | Rate of hospitalization of quarantined individuals | On average 0.125 of quarantined individuals go to hospitals. [11] |
| $d_I$ | Rate of COVID-induced death for symptomatic individuals | Median number of days that a patient from symptom group to be dead is 2 weeks. [11] |
| $d_H$ | Rate of COVID-induced death for isolated individuals | Approximately 10% of non-ICU and ICU admissions will die because of COVID-19. [11] |

Note: This table summarized the parameter's quantities that I use to run the best fit SEQIHR model.

I ran the system of equations of the SEQIHR model using the above parameters and got the Figure 5, which surprisingly fit to the 2020 daily death data of the United States. It should be noted that Atkeson (2021b) used a SEIHR model and got the similar plot.

*Figure 5. COVID-19 Daily Death Toll (Model (blue) and Data (red))*

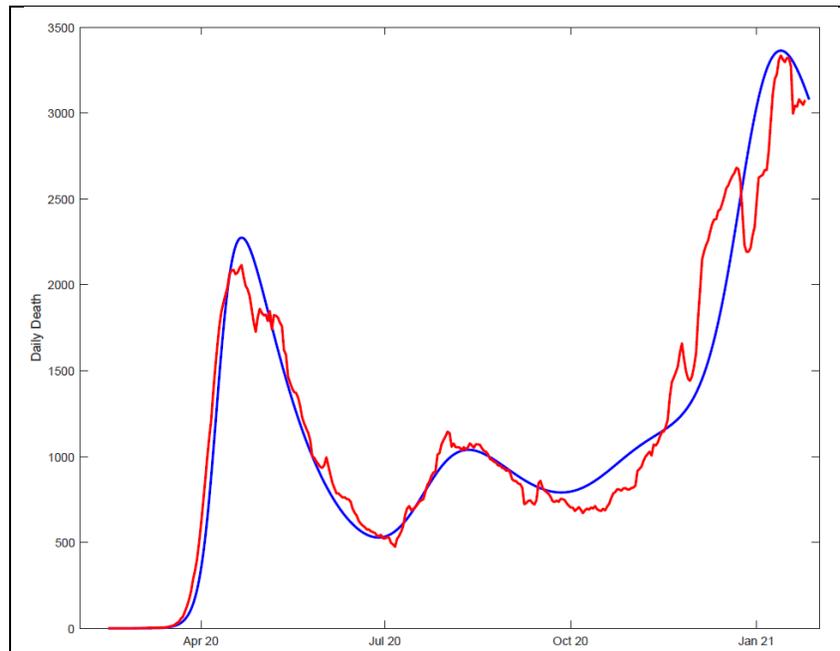

Note: The figure plots the real data (7-Day moving average) and the estimated SEQIHR model based on the Table 2 parameter's quantities. As it is clear, the estimated SEQIHR model (blue line) is a good fit to the real data.

Since an adjusted SEQIHR model did a great job to fit the data, in the following section, I used the above fitted model and its parameters' quantities to find economic effect of COVID-19 in the case that there exists heterogeneous population.

## 3.4  Macro Economics of COVID-19

There are several papers which evaluate the effectiveness on COVID-controlling policies (see Gupta et al. (2020); Dave et al. (2021);Brzezinski et al. (2020); Hsiang et al. (2020); Friedson et al. (2020)). Beside the effectiveness of COVID-controlling policies, one feature which is same among them is that they implemented uniformly. In fact, they treat all people equally and ignore the heterogeneity characteristic of population. Form other hand, one of the main issues of compartmental models is that they do not consider heterogeneity among a population. In a pandemic like COVID-19, the severity of illness and mortality rate are different for different age groups and excluding this crucial characteristic may lead to inaccurate conclusions. Experience with COVID-19 shows how good analysis could help to reduce the deadly effect of a pandemic. By analyzing and implementing good strategies, governments could decrease mortality rate and the negative effect of a pandemic on the economy. I aim to analyze the economic effect of COVID-19 by using a multi risk SEQIHR model. As shown in the previous section, the SEQIHR model is a good fit to the data in that it can capture the nonlinear features of the US daily COVID-induced death toll. Therefore, in the following paragraphs I use this model in the case that we have heterogenous compartments, susceptible, infected, hospitalized, quarantined, and recovered. Based on CDC categorization, I considered the heterogeneous characteristics of these compartments for people who are between 18 and 39 years old (young), between 40 and 64 years old (middle-age) and more than 65 years old (old). I excluded the group with those younger than 17 years old, because the mortality rate for them is very low. In the following I consider multi-risk SEQIHR model to capture the heterogeneity feature of COVID-19.

## 3.5  Multi-risk SEQIHR (MR- SEQIHR) model
In order to evaluate the economic effect of COVID-19 I use the same methodology introduced by Acemoglu, Chernozhukov, et al. (2020) .However, there are several main differences between their approach and mine.

- As it shown in the previous part the SEQIHR model did a good job to fit the real daily death toll, hence I use the multi-risk SEQIHR model, while Acemoglu, Chernozhukov, et al. (2020) used multi-risk SIR model.
- Acemoglu et al. 2020 published the paper in May 2020, because of that they had limited data and inaccurate parameters' quantities. However, I have the COVID-19 related data up until December 2021, so that I have more accurate parameters' quantities. Because of that, as it shown in the results part, my results are very close to the real situation compared to Acemoglu, Chernozhukov, et al. (2020).
- Moreover, there are some incorrect assumptions in the Acemoglu, Chernozhukov, et al. (2020) which lead to unvalidated results and somehow far from the truth. For instance, in their paper they mentioned non-ICU patient do not die, while in the reality it is not true. Also, they asserted the proportion of ICU and non-ICU patients do not change over time, though the data show this proportion changed over time. Moreover, they argued that recovered individuals are immune against COVID-19 infection, whilst based on the COVID-19 data there is a possibility that recovered people got infected again. Furthermore, they considered stochastic vaccine arrival, however, we know that the public COVID-19 vaccination was started at late December 2020.

Based on the adequate data and more accurate parameters' quantities and correct assumptions I ran the MR-SEQIHR model. As it is shown in the Figure 6, I divided each compartment of the SEQIHR model which presented in the previous part based on three age groups: 18 and 39 years old (young), between 40 and 64 years old (middle-age) and more than 65 years old (old). Hence, the total population formula denotes in the equation ( 26 ).

$$N_i(t) = S_i(t) + E_i(t) + I_i(t) + Q_i(t) + H_i(t) + R_i(t) \qquad (26)$$

Where $i$ indicates different age groups and t is continuous time; $t \in [0, \infty)$. Also, the total population is normalized to unity, hence $\sum_i N_i = 1$.

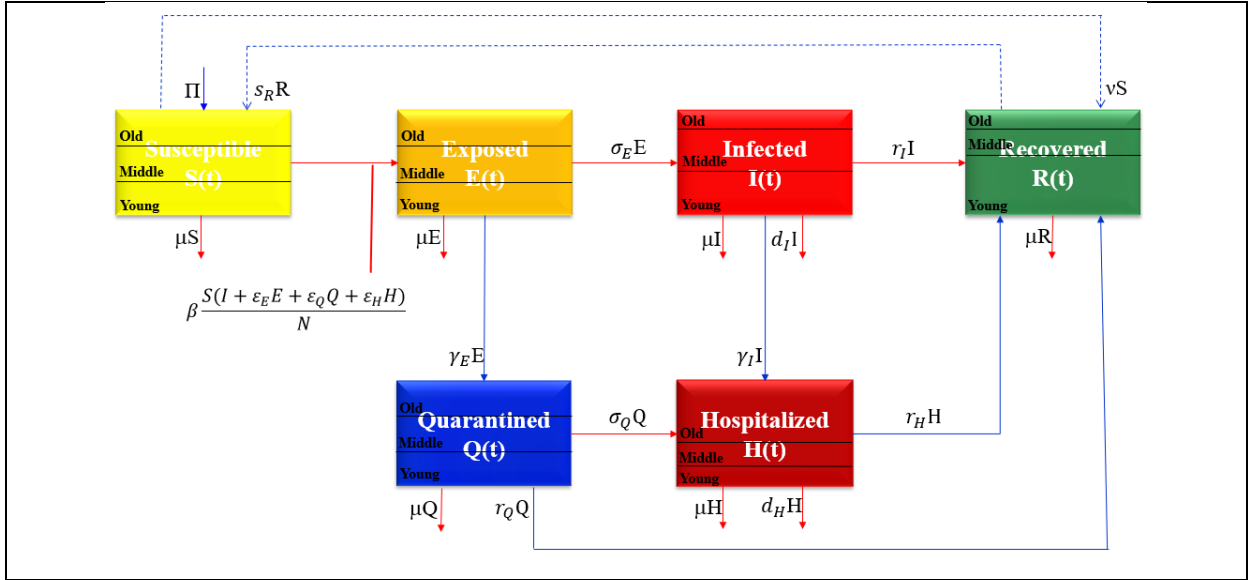

Figure 6. The Multi-risk SEQIHR (MR-SEQIHR) Model

Note: By considering heterogenous population and the SEQIHR model assumptions I depict the multi-risk SEQIHR plot which divided each compartment into three categories: 18 and 39 years old (young), between 40 and 64 years old (middle-age) and more than 65 years old (old).

In addition to the relationship between the compartments described previously, different age groups have interactions within each compartment.

In order to evaluate the economic effect of COVID-19, beside the SEQIHR Model's parameters that I mentioned in the previous section, I use the following parameters to run MR-SEQIHR model.

- $w_i$ denotes the workers' production of age group i in the absence of lockdown, otherwise the production is zero. w can represent the wage of employees too.
- $L_i(t)$ indicates the fraction of age group i that is in lockdown. $L_i(t) \in [0,1]$, where $L_i(t) = 0$ indicates no lockdown, and $L_i(t) = 1$ indicates full lockdown.
- $\chi_i$ designates cost of death in age group i other than production lost when employees are dead because of COVID-19.
- $\kappa_i$ refers the share of recovered individuals in age group i who allowed to work.
- $r$ indicates the interest rate in order to calculate the present value of monetary and non-monetary costs of COVID-induced death.

Susceptible (S), exposed (E), infected (I), quarantined (Q), isolated/hospitalized (H), and recovered/dead (R) dynamics are similar to the equations ( 1 ) to ( 6 ), Which explained in the previous section. I analyze the economic effect of COVID-19 through employment; hence I consider some assumptions in order to get the employment equation. These assumptions are as follows.

### 3.6 MR-SEQIHR Model Assumptions

- The fraction of susceptible individuals of each age group who allow to work is $(1 - \mu - L_i(t))$.
- The fraction of exposed individuals of each age group who allow to work is $(1 - \mu - L_i(t))$. In addition, the fraction of exposed individuals of each age groups go to quarantined and infected compartments with rates $\gamma_E$, and $\sigma_E$ respectively, so that they are not allowed to work.
- The fraction of infected individuals of each age group who allow to work is $(1 - \mu - L_i(t))$. Moreover, the fraction of infected individuals of each age groups go to hospitalized compartment with rate $\gamma_I$ or die because of COVID-19 with rate $d_I$, so that they are not allowed to work.
- Based on the input and output of quarantined compartment, the fraction of quarantined individuals of each age group who cannot work is $(1 - r_Q)$.
- Based on the input and output of hospitalized compartment, the fraction of hospitalized individuals of each age group who cannot work is $(1 - r_H)$.
- The fraction of recovered individuals of each age group who allow to work is $(1 - \mu - L_i(t))$. In addition, $\kappa_i$ fraction of recovered individuals in each age group are allowed to work.

Therefore, based on the above assumptions we could write the employments equation as follows:

$$EMP_i(t) = (1 - \mu - L_i(t))(S_i(t) + E_i(t) + I_i(t) + R_i(t)) - (\gamma_E + \sigma_E)E_i(t) - (\gamma_I + d_I)I_i(t) - (1 - r_Q)Q_i(t) - (1 - r_H)H_i(t) + \kappa_i R_i(t) \qquad (27)$$

Furthermore, the planner's objective is controlling lockdown, $\{L_i(t)\}_i$, for $t \in [0, T)$, in order to minimize the expected present value of social costs. Then, we could write the objective function as follows:

$$\int_0^\infty e^{rt} \sum_i (w_i(N_i - EMP_i(t)) + \chi_i(d_I I_i(t) + d_H H_i(t))) dt \qquad (28)$$

Note that $w_i(N_i - EMP_i(t))$ denotes monetary costs of death and $\chi_i(d_I I_i(t) + d_H H_i(t))$ represents non-monetary costs of death. I consider the period before arriving vaccine, hence by substituting equation ( 27 ) in equation ( 28 ), we can rewrite the objective function as follows:

$$\int_0^\infty e^{rt} \sum_i (w_i(N_i - ((1 - \mu - L_i(t))(S_i(t) + E_i(t) + I_i(t) + R_i(t)) - (\gamma_E + \sigma_E)E_i(t) - (\gamma_I + d_I)I_i(t) - (1 - r_Q)Q_i(t) - (1 - r_H)H_i(t) + \kappa_i R_i(t))) + \hat{\chi}_i(d_I I_i(t) + d_H H_i(t)))dt \quad (29)$$

Where $\hat{\chi}_i = \frac{w_i}{r} + \chi_i$ indicates the total cost of death.

### 3.7 MR- SEQIHR Model calibration

Since SEQIHR did a good job to fit the real data, the parameters' quantities which I used to run SEQIHR in the previous section are the same. Other required quantities for running MR- SEQIHR are as follows:

- Based on the Bureau of Labor Statistics (BLS) report of 2020 [14], among the population who are older than 18 years, the fraction who are young (18-39 years old), $N_y = 0.542$, the fraction who are middle-aged (between 40 and 64 years old), $N_m = 0.246$, and the fraction who are old (more than 65 years old), $N_o = 0.211$.

- I assume that the old people produce nothing, $w_o = 0$, and young and middle-aged workers' production equal 1, $w_y = w_m = 1$. [15]

- Since $L_i(t)$ represents the severity of lockdown, I assume $\overline{L_o} = 1$, and $\overline{L_y} = \overline{L_m} = 0.7$. Moreover, in the case of uniform policy I assume $\overline{L} = 0.7$. (Alvarez et al).

- Based on the CDC report, the COVID-induced mortality rate of young is 0.000315, the mortality of middle-age is 0.00132, and the mortality rate of old is 0.0030. [16]

- As in [Acemoglu, Chernozhukov, et al. (2020)](), I set non-monetary cost of death, $\chi_i = \chi - \frac{w_i}{r}e^{-r\Delta_i}$. Where $\Delta_i$ denotes remaining work time, so that $\Delta_{Ac} = 15 \times 365$ and $\Delta_{Re} = 0$. Also, daily interest rate, $r = \frac{0.01}{365}$. Moreover, I consider $\chi = 20$ in the baseline model, which means non-monetary cost of death is 20 times the annual economic contribution of a representative worker.

- [Acemoglu, Chernozhukov, et al. (2020)](#) consider stochastic vaccine arrival, while we know that the first dose of vaccine was available for public at late December 2020. Because of that I just consider the situation that we have deterministic vaccine arrival. In the baseline model, I consider T=365, which implies it takes around a year the first doses of vaccine to be available publicly.

# 4. Results

Acemoglu, Chernozhukov, et al. (2020) used pareto frontier to show the trade-off between COVID-19 induced life loss and GDP loss as an indicator of economic loss of the United States. They show this relationship in the Figure 7. As it is clear after a certain point there is an ascending trend in the pareto frontier which indicates more GDP and lives lost in the absence of any COVID-19 controlling policies. In addition, the optimal targeted curve is closer to the "bliss point", the origin, than the uniform curve.

*Figure 7.Pareto Frontier of COVID-controlling policies*

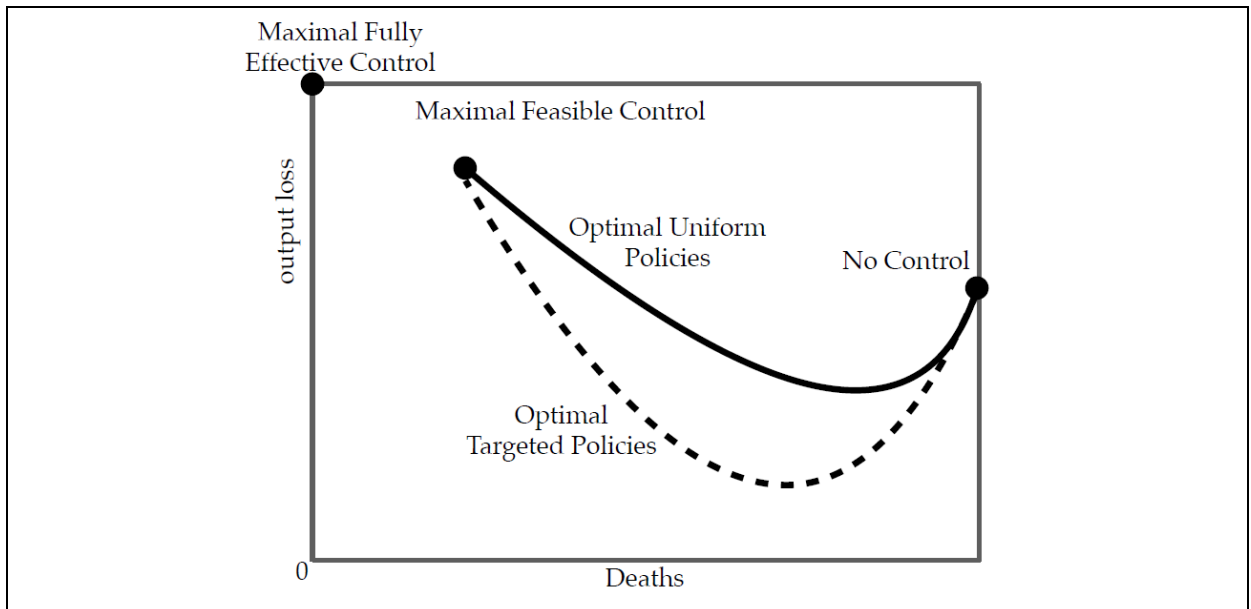

Note: This figure represents the pareto frontier of optimal uniform and targeted COVID- controlling policies. As it is clear the optimal targeted policies which treated distinct age groups differently could reduce the output loss and death rate much better than uniform policies which treat population uniformly, because of that the optimal targeted curve is closer to the "bliss point", the origin, than the uniform curve.Acemoglu, Chernozhukov, et al. (2020)

They argue that when we have uniform policy which means all the population treated equally for implementing lockdown policies, the death rate and GDP loss will be 1.83 percent and 23.4 percent respectively at the end of 2020. However, targeted policies and even simi targeted policies which treated different age groups differently and impose strict lockdown on old individuals could reduce the fatality rate and GDP loss compared to uniform policies. Specifically, the optimal targeted policies that contains strict lockdown on oldest group can reduce the fatality rate and GDP loss to 1 percent and 12.8 percent respectively compared to uniform lockdown.

However, based on the World bank data,[17] the GDP growth rate of united states in 2020 is -3.5 percent, while this indicator was 2.2 percent in 2019. Comparing actual statistics with Acemoglu, Chernozhukov, et al. (2020) results indicate a significant difference between them. Even, optimal targeted policies which computed 12.8 percent decrease in GDP is far from the reality. Moreover, based of the National Center for Health Statistics (NCHS) report,[18] which mentioned previously, the COVID-19 induced-death rate is 0.1 percent which is far different from Acemoglu, Chernozhukov, et al. (2020) results. As previously discussed, wrong assumptions, inaccurate parameters, and limited data may lead to these incorrect results.

In order to get the precise results, I use the actual parameters and good fitted SEQIHR model. Figure 8 represents pareto frontier of GDP loss and death rate under distinct COVID-controlling policies for different quantities of $\chi$. Like figure 7 the origin, where there is zero death rate and GDP loss, indicates bliss point. The convex shape of the frontiers indicates the decreasing returns to scale of getting one goal in the expense of other. The red point refers the exact condition of US during 2020 where the GDP loss is 3.5 percent and death rate is 0.1 percent based on the real data. In addition, the blue dotted line represents the uniform COVID-controlling policies which treated all people uniformly. Moreover, the green dotted line represents the targeted COVID-controlling policies which treated each age group differently.

*Figure 8.Pareto Frontier of Targeted and Uniform COVID-controlling Policies*

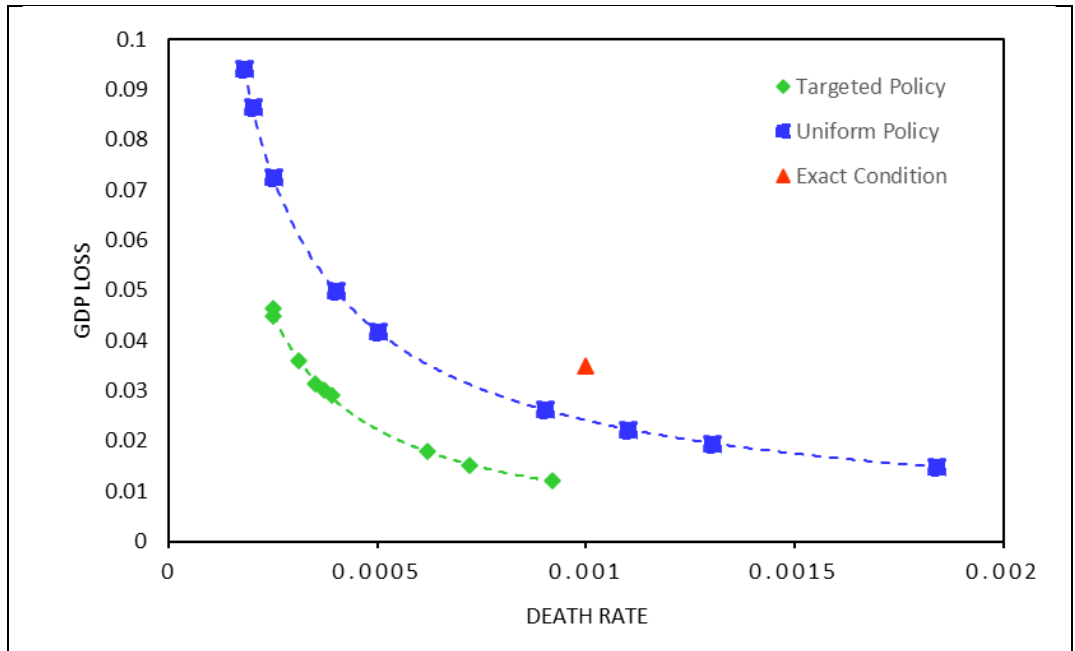

Note: The figure plots the pareto frontier of targeted and uniform COVID-controlling policies plus the exact condition of death rate and GDP loss based on the real data. The blue dotted line represents the uniform COVID-controlling policies, the green dotted line depicts the targeted COVID-controlling policies, and the red dot refers to exact condition of the US based on the real data.

As it is clear from Figure 8, the exact condition of US is close to optimal uniform policies. Also, GDP maximizing outcome occurred in the southeast part of each frontier. Then, at the GDP maximizing point of the optimal uniform policy frontier, the GDP loss is 2 percent and death rate is 0.18 percent. However, at the GDP maximizing point of the optimal targeted policy frontier, the GDP loss is 1 percent and death rate is 0.09 percent. More importantly, with the same level of GDP loss of exact condition, 3.5 percent, optimal targeted policy could reduce the death rate to 0.03 percent. Declining death toll by 0.07 percent means for the same percent of GDP loss if we implement optimal targeted policy, we could save 2/3 lives that is dead because of COVID-19. Nearly, 378,000 persons dead because of COVID-19 during 2020, therefore reducing death rate to 0.03 percent mean saving around 280,000 lives, which is huge.

# 5. Conclusion

All in all, the adjusted SEQIHR did a great job to show the dynamic of COVID-19 over time, specifically it is good fit to the COVID-induced daily death data of United States in a way that it could capture the nonlinearities of the data very well. Because of that I used this model to estimate the economic impact of COVID-19 in the United States. Since risk of hospitalization, infection and death are different for various age groups, I used multi-risk SEQIHR model to evaluate lockdown policies. I considered the heterogeneous feature of COVID-19 by dividing each compartment of SEQIHR model into three categories: individuals who are between 18 and 39 years old (young), persons who are between 40 and 64 years old (middle-age) and people who are more than 65 years old (old). Since the mortality rate for those younger than 17 years old is very low I excluded them, then adding several parameters to the SEQIHR model and ran the model. Based on the actual data, the GDP loss and death rate for the United States during 2020 were 3.5 and 0.1 percent. However, the paper's results show that optimal targeted policy which imposed more severe restrictions on old people could reduce mortality rate to 0.03 percent with the same level of GDP loss of actual data, 3.5 percent. Compared to actual data, the targeted policy could save 280,000 lives.

## 6. Notes

1. The IMF report available at https://www.imf.org/en/Publications/WEO/Issues/2021/03/23/world-economic-outlook-april-2021
2. Gating criteria are the data-driven conditions each region or state should satisfy before proceeding to a phased opening.
3. The COVID-19 timeline available at https://www.cdc.gov/museum/timeline/covid19.html and https://www.nytimes.com/article/coronavirus-timeline.html
4. The information regarding risk for COVID-19 infection, hospitalization, and death by age group available at https://www.cdc.gov/coronavirus/2019-ncov/covid-data/investigations-discovery/hospitalization-death-by-age.html
5. The data related to the number of COVID-19 positive cases by age group available at https://covid.cdc.gov/covid-data-tracker/#demographics
6. The data related to the number of COVID-induced death by age group available at https://covid.cdc.gov/covid-data-tracker/#demographics
7. Based on the National Center for Health Statistics (NCHS) report, the provisional general fertility rate (GFR) for the United States in 2020 was 55.8 births per 1,000 women aged 15–44. The NCHS report available at https://www.cdc.gov/nchs/nvss/vsrr/natality-dashboard.htm
8. In 2020, approximately 3,358,814 deaths (828.7 deaths per 100,000 population) recorded in the United States. During 2020, COVID-19 was the third leading cause of death with 377,883 deaths (91.5 per 100,000 population). Therefore, the natural death rate for the United States is 7.37 per 1000 people in 2020. The mortality report available at https://www.cdc.gov/mmwr/volumes/70/wr/mm7014e1.htm
9. The data regarding the COVID-19 vaccination in the United States available at https://ourworldindata.org/covid-vaccinations?country=USA

10. The data regarding COVID-19 vaccine breakthrough infection available at https://www.hopkinsmedicine.org/health/conditions-and-diseases/coronavirus/breakthrough-infections-coronavirus-after-vaccination

11. Based on CDC's COVID-19 pandemic planning scenarios the mean time that infectious person should be quarantine is 2 weeks. The median number of days from symptom group to hospitalized group is 5 days. 98% of COVID-19 infectious individual recovered. 80% of patients hospitalized with COVID-19, and 60% of those admitted to ICUs, survive. On average a COVID patient needs 2 weeks to be recover. Mean time from exposure group to symptom group is 6 days. On average 0.125 of quarantined individuals go to hospitals. Median number of days that a patient from symptom group to be dead is 2 weeks. Approximately 10% of non-ICU and ICU admissions will die because of COVID-19. These scenarios are available at https://www.cdc.gov/coronavirus/2019-ncov/hcp/planning-scenarios.html

14. The Bureau of Labor Statistics (BLS) report of 2020 available at https://www.bls.gov/cps/demographics.htm

15. In fact, based on the actual data which release by BLS, around 20 percent of old individuals (more than 65 years old) are employed, while around 78 percent of young people (18-39 years old) and 68 percent of middle-aged people (between 40 and 64 years old) are employed. In the robustness part I check this issue by considering different quantities of w for different age groups. Regarding information available at https://www.bls.gov/cps/demographics.htm

16. The COVID-19 mortality rate based on age group available at https://www.cdc.gov/nchs/nvss/vsrr/covid_weekly/index.htm

17. The World bank data available at https://data.worldbank.org/.

18. The National Center for Health Statistics available at https://nchstats.com/

# 7. Reference

Acemoglu, Daron, Victor Chernozhukov, Iván Werning, and Michael D Whinston. 2020. *A multi-risk SIR model with optimally targeted lockdown*. Vol. 2020: National Bureau of Economic Research Cambridge, MA.

Acemoglu, Daron, Alireza Fallah, A Giometto, D Huttenlocher, A Ozdaglar, Francesca Parise, and Sarath Pattathil. 2021. "Optimal adaptive testing for epidemic control: combining molecular and serology tests." *arXiv preprint arXiv:2101.00773*.

Acemoglu, Daron, Ali Makhdoumi, Azarakhsh Malekian, and Asuman Ozdaglar. 2020. Testing, voluntary social distancing and the spread of an infection. national bureau of economic Research.

Arias, Jonas E, Jesús Fernández-Villaverde, Juan Rubio Ramírez, and Minchul Shin. 2021. Bayesian Estimation of Epidemiological Models: Methods, Causality, and Policy Trade-Offs. National Bureau of Economic Research.

Arnon, Alexander, John Ricco, and Kent Smetters. 2020. "Epidemiological and economic effects of lockdown." *Brookings Papers on Economic Activity*.

Atkeson, Andrew. 2020a. How deadly is COVID-19? Understanding the difficulties with estimation of its fatality rate. National Bureau of Economic Research.

Atkeson, Andrew. 2020b. What will be the economic impact of COVID-19 in the US? Rough estimates of disease scenarios. National Bureau of Economic Research.

Atkeson, Andrew. 2021a. Behavior and the Dynamics of Epidemics. National Bureau of Economic Research.

Atkeson, Andrew. 2021b. A parsimonious behavioral SEIR model of the 2020 COVID epidemic in the United States and the United Kingdom. National Bureau of Economic Research.

Atkeson, Andrew, Michael C Droste, Michael Mina, and James H Stock. 2020. Economic benefits of covid-19 screening tests. National Bureau of Economic Research.

Atkeson, Andrew G, Karen Kopecky, and Tao Zha. 2021. "Behavior and the Transmission of COVID-19." AEA Papers and Proceedings.

Atkeson, Andrew, Karen Kopecky, and Tao Zha. 2020a. Estimating and forecasting disease scenarios for COVID-19 with an SIR model. National Bureau of Economic Research.

Atkeson, Andrew, Karen Kopecky, and Tao Zha. 2020b. Four stylized facts about COVID-19. National Bureau of Economic Research.

Avery, Christopher, William Bossert, Adam Clark, Glenn Ellison, and Sara Fisher Ellison. 2020. "Policy implications of models of the spread of coronavirus: Perspectives and opportunities for economists."

Baker, Scott R, Nicholas Bloom, Steven J Davis, and Stephen J Terry. 2020. Covid-induced economic uncertainty. National Bureau of Economic Research.

Baqaee, David, and Emmanuel Farhi. 2020. Nonlinear production networks with an application to the covid-19 crisis. National Bureau of Economic Research.

Baqaee, David, and Emmanuel Farhi. 2021. "Keynesian Production Networks and the Covid-19 Crisis: A Simple Benchmark." AEA Papers and Proceedings.

Baqaee, David, Emmanuel Farhi, Michael J Mina, and James H Stock. 2020. Reopening scenarios. National Bureau of Economic Research.

Barua, Suborna. 2020. "Understanding Coronanomics: The economic implications of the coronavirus (COVID-19) pandemic." *Available at SSRN 3566477*.

Berger, David W, Kyle F Herkenhoff, and Simon Mongey. 2020. An seir infectious disease model with testing and conditional quarantine. National Bureau of Economic Research.

Bertozzi, Andrea L, Elisa Franco, George Mohler, Martin B Short, and Daniel Sledge. 2020. "The challenges of modeling and forecasting the spread of COVID-19." *Proceedings of the National Academy of Sciences* 117 (29):16732-16738.

Blackwood, Julie C, and Lauren M Childs. 2018. "An introduction to compartmental modeling for the budding infectious disease modeler."

Bloom, David E, Michael Kuhn, and Klaus Prettner. 2020. Modern infectious diseases: Macroeconomic impacts and policy responses. National Bureau of Economic Research.

Bloom, Nicholas, Robert S Fletcher, and Ethan Yeh. 2021. The impact of COVID-19 on US firms. National Bureau of Economic Research.

Boppart, Timo, Karl Harmenberg, John Hassler, Per Krusell, and Jonna Olsson. 2020. Integrated epi-econ assessment. National Bureau of Economic Research.

Bourne, Ryan A. 2021. *Economics in One Virus: An Introduction to Economic Reasoning through COVID-19*: Cato Institute.

Brauer, Fred, Carlos Castillo-Chavez, and Zhilan Feng. 2019. *Mathematical models in epidemiology*. Vol. 32: Springer.

Brinca, Pedro, Joao B Duarte, and Miguel Faria-e-Castro. 2020. "Measuring sectoral supply and demand shocks during COVID-19." *Frb st. louis working paper* (2020-011).

Brodeur, Abel, David Gray, Anik Islam, and Suraiya Bhuiyan. 2021. "A literature review of the economics of COVID-19." *Journal of Economic Surveys* 35 (4):1007-1044.

Brotherhood, Luiz, Philipp Kircher, Cezar Santos, and Michele Tertilt. 2020. "An economic model of the Covid-19 pandemic with young and old agents: Behavior, testing and policies." *University of Bonn and University of Mannheim, Germany, Discussion Paper No* 175.

Brzezinski, Adam, Guido Deiana, Valentin Kecht, and David Van Dijcke. 2020. "The covid-19 pandemic: government vs. community action across the united states." *Covid Economics: Vetted and Real-Time Papers* 7:115-156.

Cajner, Tomaz, Leland D Crane, Ryan A Decker, John Grigsby, Adrian Hamins-Puertolas, Erik Hurst, Christopher Kurz, and Ahu Yildirmaz. 2020. The US labor market during the beginning of the pandemic recession. National Bureau of Economic Research.

Chen, Simiao, Klaus Prettner, Michael Kuhn, and David E Bloom. 2021. "The economic burden of COVID-19 in the United States: Estimates and projections under an infection-based herd immunity approach." *The Journal of the Economics of Ageing*:100328.

Chernozhukov, Victor, Hiroyuki Kasahara, and Paul Schrimpf. 2021. "Causal impact of masks, policies, behavior on early covid-19 pandemic in the US." *Journal of econometrics* 220 (1):23-62.

COVID, IHME, Forecasting Team, and Simon I Hay. 2020. "COVID-19 scenarios for the United States." *medRxiv*.

Crossley, Thomas F, Paul Fisher, and Hamish Low. 2021. "The heterogeneous and regressive consequences of COVID-19: Evidence from high quality panel data." *Journal of public economics* 193:104334.

Dave, Dhaval, Andrew I Friedson, Kyutaro Matsuzawa, and Joseph J Sabia. 2021. "When do shelter-in-place orders fight COVID-19 best? Policy heterogeneity across states and adoption time." *Economic inquiry* 59 (1):29-52.

Diekmann, Odo, Johan Andre Peter Heesterbeek, and Johan AJ Metz. 1995. "The legacy of Kermack and McKendrick." *Epidemic models: their structure and relation to data* 5:95.

Eichenbaum, Martin S, Sergio Rebelo, and Mathias Trabandt. 2021. "The macroeconomics of epidemics." *The Review of Financial Studies* 34 (11):5149-5187.

Farboodi, Maryam, Gregor Jarosch, and Robert Shimer. 2021. "Internal and external effects of social distancing in a pandemic." *Journal of Economic Theory* 196:105293.

Favero, Carlo A, Andrea Ichino, and Aldo Rustichini. 2020. "Restarting the economy while saving lives under Covid-19."

Fernández-Villaverde, Jesús, and Charles I Jones. 2020. Estimating and simulating a SIRD model of COVID-19 for many countries, states, and cities. National Bureau of Economic Research.

Friedman, Eric, John Friedman, Simon Johnson, and Adam Landsberg. 2020. "Transitioning out of the coronavirus lockdown: A framework for evaluating zone-based social distancing." *Frontiers in public health* 8:266.

Friedson, Andrew I, Drew McNichols, Joseph J Sabia, and Dhaval Dave. 2020. Did California's shelter-in-place order work? Early coronavirus-related public health effects. National Bureau of Economic Research.

Gibson, Gavin J, and Eric Renshaw. 1998. "Estimating parameters in stochastic compartmental models using Markov chain methods." *Mathematical Medicine and Biology: A Journal of the IMA* 15 (1):19-40.

GILL, NATHAN. "DETERMINISTIC AND STOCHASTIC MODELS OF INFECTIOUS DISEASE: CIRCULAR MIGRATIONS AND HIV TRANSMISSION DYNAMICS."

Glover, Andrew, Jonathan Heathcote, Dirk Krueger, and José-Víctor Ríos-Rull. 2020. Health versus wealth: On the distributional effects of controlling a pandemic. National Bureau of Economic Research.

Gollier, Christian. 2020. "Cost–benefit analysis of age-specific deconfinement strategies." *Journal of Public Economic Theory* 22 (6):1746-1771.

Guerrieri, Veronica, Guido Lorenzoni, Ludwig Straub, and Iván Werning. 2020. Macroeconomic implications of COVID-19: Can negative supply shocks cause demand shortages? : National Bureau of Economic Research.

Gumel, Abba B, Shigui Ruan, Troy Day, James Watmough, Fred Brauer, P Van den Driessche, Dave Gabrielson, Chris Bowman, Murray E Alexander, and Sten Ardal. 2004. "Modelling strategies for controlling SARS outbreaks." *Proceedings of the Royal Society of London. Series B: Biological Sciences* 271 (1554):2223-2232.

Gupta, Sumedha, Thuy D Nguyen, Felipe Lozano Rojas, Shyam Raman, Byungkyu Lee, Ana Bento, Kosali I Simon, and Coady Wing. 2020. Tracking public and private responses to the COVID-19 epidemic: evidence from state and local government actions. National Bureau of Economic Research.

Hall, Robert E, Charles I Jones, and Peter J Klenow. 2020. Trading off consumption and covid-19 deaths. National Bureau of Economic Research.

Heesterbeek, JAP, and MG Roberts. 2007. "The type-reproduction number T in models for infectious disease control." *Mathematical biosciences* 206 (1):3-10.

Hethcote, Herbert W. 2000. "The mathematics of infectious diseases." *SIAM review* 42 (4):599-653.

Hsiang, Solomon, Daniel Allen, Sébastien Annan-Phan, Kendon Bell, Ian Bolliger, Trinetta Chong, Hannah Druckenmiller, Luna Yue Huang, Andrew Hultgren, and Emma Krasovich. 2020. "The effect of large-scale anti-contagion policies on the COVID-19 pandemic." *Nature* 584 (7820):262-267.

Kaplan, Greg, Benjamin Moll, and Giovanni L Violante. 2020. The great lockdown and the big stimulus: Tracing the pandemic possibility frontier for the US. National Bureau of Economic Research.

Kermack, William Ogilvy, and Anderson G McKendrick. 1927. "A contribution to the mathematical theory of epidemics." *Proceedings of the royal society of london. Series A, Containing papers of a mathematical and physical character* 115 (772):700-721.

Kermack, William Ogilvy, and Anderson G McKendrick. 1932. "Contributions to the mathematical theory of epidemics. II.—The problem of endemicity." *Proceedings of the Royal Society of London. Series A, containing papers of a mathematical and physical character* 138 (834):55-83.

Kermack, William Ogilvy, and Anderson G McKendrick. 1933. "Contributions to the mathematical theory of epidemics. III.—Further studies of the problem of endemicity." *Proceedings of the Royal Society of London. Series A, Containing Papers of a Mathematical and Physical Character* 141 (843):94-122.

La Torre, Davide, Tufail Malik, and Simone Marsiglio. 2020. "Optimal control of prevention and treatment in a basic macroeconomic–epidemiological model." *Mathematical Social Sciences* 108:100-108.

McKibbin, Warwick, and Roshen Fernando. 2020. "The economic impact of COVID-19." *Economics in the Time of COVID-19* 45 (10.1162).

McKibbin, Warwick, and Roshen Fernando. 2021. "The global macroeconomic impacts of COVID-19: Seven scenarios." *Asian Economic Papers* 20 (2):1-30.

Rahmandad, Hazhir, Tse Yang Lim, and John Sterman. 2021. "Behavioral dynamics of COVID-19: Estimating under-reporting, multiple waves, and adherence fatigue across 92 nations." *Rahmandad, H., Lim, TY., Sterman, J., Behavioral Dynamics of COVID-19: Estimating Under-Reporting, Multiple Waves, and Adherence Fatigue Across* 92.

Rampini, Adriano A. 2020. Sequential lifting of COVID-19 interventions with population heterogeneity. National Bureau of Economic Research.

Roda, Weston C, Marie B Varughese, Donglin Han, and Michael Y Li. 2020. "Why is it difficult to accurately predict the COVID-19 epidemic?" *Infectious Disease Modelling* 5:271-281.

Sargent, Thomas, and John Stachurski. 2015. "Quantitative economics with Julia."

Silal, Sheetal Prakash, Francesca Little, Karen I Barnes, and Lisa Jane White. 2016. "Sensitivity to model structure: a comparison of compartmental models in epidemiology." *Health Systems* 5 (3):178-191.

Siriprapaiwan, Supatcha, Elvin J Moore, and Sanoe Koonprasert. 2018. "Generalized reproduction numbers, sensitivity analysis and critical immunity levels of an SEQIJR disease model with immunization and varying total population size." *Mathematics and computers in simulation* 146:70-89.

Stock, James H. 2020. Data gaps and the policy response to the novel coronavirus. National Bureau of Economic Research.

Wang, Shuo, Xian Yang, Ling Li, Philip Nadler, Rossella Arcucci, Yuan Huang, Zhongzhao Teng, and Yike Guo. 2020. "A bayesian updating scheme for pandemics: estimating the infection dynamics of covid-19." *IEEE Computational Intelligence Magazine* 15 (4):23-33.

Willebrand, Kathryn. 2021. "Transmission Dynamics of COVID-19 under Differential Levels of Masking and Vaccination." Yale University.

Worden, Lee, and Travis C Porco. 2017. "Products of Compartmental Models in Epidemiology." *Computational and mathematical methods in medicine* 2017.

Wu, Desheng Dash, and David L Olson. 2020. *Pandemic risk management in operations and finance*: Springer.

Yan, Xiefei, and Yun Zou. 2006. "Optimal quarantine and isolation control in SEQIJR SARS model." 2006 9th International Conference on Control, Automation, Robotics and Vision.

Zhang, Chunyan, Songhua Xu, Zongfang Li, and Shunxu Hu. 2021. "Understanding Concerns, Sentiments, and Disparities Among Population Groups During the COVID-19 Pandemic Via Twitter Data Mining: Large-scale Cross-sectional Study." *Journal of medical Internet research* 23 (3):e26482.